\documentclass[a4paper,twocolumn,showpacs]{revtex4}

\usepackage{amsmath,amsfonts,amssymb}
\usepackage{graphicx}	
\usepackage{bbm}

\newcommand{\ie}{i.e.\ }

\newcommand{\myNeg}{\negthickspace\negthickspace}

\newcommand{\bra}[1]{\langle #1|\,}
\newcommand{\ket}[1]{\,|#1 \rangle}

\newcommand{\ketbra}[2]{\,|#1 \rangle\negmedspace\langle #2|\,}

\newcommand{\eps}{\varepsilon}
\newcommand{\id}{\mathbbm{1}}
\newcommand{\cA}{\mathcal{A}}
\newcommand{\cB}{\mathcal{B}}
\newcommand{\cC}{\mathcal{C}}
\newcommand{\cL}{\mathcal{L}}
\newcommand{\cN}{\mathcal{N}}
\newcommand{\cR}{\mathcal{R}}
\newcommand{\cT}{\mathcal{T}}
\newcommand{\GLL}{\Gamma_{\cL\cL}}
\newcommand{\GLR}{\Gamma_{\cL\cR}}

\newcommand{\GTL}{\Gamma_{\cT\cL}}
\newcommand{\GTB}{\Gamma_{\cT\cB}}
\newcommand{\cs}{\ket{C}}
\newcommand{\rs}{\ket{\psi}_{AB}}

\DeclareMathOperator{\prob}{prob}

\newcommand{\ps}{2.27} 
\newcommand{\dps}{0.03}
\newcommand{\es}{3.86}
\newcommand{\fx}{0.699}
\newcommand{\dfx}{0.003}

\renewenvironment{itemize}{%
	\begin{list}{$\bullet\,$}{\leftmargin=1em}}{%
	\end{list}}

\begin{document}

\title{Fidelity threshold for long-range entanglement in quantum networks}
\author{S\'ebastien Perseguers}
\email{sebastien.perseguers@mpq.mpg.de}
\affiliation{Max-Planck--Institut f\"ur Quantenoptik, Hans-Kopfermann-Strasse 1, D-85748 Garching, Germany}
\date{\today}

\begin{abstract}
A strategy to generate long-range entanglement in noisy quantum networks is presented.
We consider a cubic lattice whose bonds are partially entangled mixed states of two
qubits, and where quantum operations can be applied perfectly at the nodes. In contrast
to protocols designed for one-~or two-dimensional regular lattices, we find that entanglement
can be created between arbitrarily distant qubits if the fidelity of the bonds is
higher than a critical value, independent of the system size. Therefore, we show that
a constant overhead of local resources, together with connections of finite fidelity,
is sufficient to achieve long-distance quantum communication in noisy networks.
\end{abstract}

\maketitle

\section{Introduction}
Quantum networks play a major role in quantum information processing \cite{K08},
as in distributed quantum computation or in quantum communication \cite{CEHM99, BBC+93}.
In fact they naturally describe the situation where neighboring stations (nodes)
share partially entangled states of qubits (noisy links).
One of the main tasks of quantum information processing
is then to design protocols that establish entanglement between any pair
of nodes, regardless of their distance in the network.

Quantum repeaters offer
a first solution to this question: one can efficiently entangle the two extremities
of a one-dimensional lattice of size $N$ by iterating purification steps and entanglement
swappings \cite{DBCZ99, DLCZ01}. This strategy needs $\mathcal{O}(\log\,N)$ qubits at each
node and runs in a time that scales as $\mathcal{O}(\text{poly}\,N)$. Though
being very promising, their realization raises some technical problems, such as the need for reliable quantum
memories \cite{HKBD07}, or the difficulty in manipulating many qubits per station.
The latter difficulty is surmounted in \cite{CTSL05}, where only
a constant number of qubits is required at each station. Various protocols
improving the rate of long-distance quantum communication have been proposed
over the past few years (see \cite{SSRG09} and references therein), but either their
time scaling remains polynomial in $N$ or they are based on rather complicated
quantum error correcting codes \cite{JTN+09}.

Motivated by the discovery of powerful protocols in the case of two-dimensional
pure-state networks \cite{ACL07}, another scheme for entanglement generation over
long distance in noisy networks was presented in \cite{PJS+08}. It
exploits the higher connectivity of the nodes to gain information on the
errors introduced by the noisy teleportations. This leads to a ``one-shot'' protocol
where elementary entangled pairs are used only once, which thus relaxes the
requirement of efficient quantum memories; see \cite{FWL+09} for the
latest quantum communication protocol in square lattices. However, the overhead
of local resources in these two-dimensional systems still slightly increases with
$N$ (logarithmic dependence).

In this work, we show that entanglement generation over arbitrarily long distance and using
the minimum amount of resources (constant number of qubits per node
and quantum operations executed in a constant time)
can be achieved in three-dimensional lattices. For this
result to hold, the fidelity $F$ of the elementary links has to be larger than
a threshold $F^*$. We first provide an analytical upper bound on this value and
then present a numerical estimate based on Monte Carlo simulations. 

\section{Description of the model}
We consider a cubic network that consists of $N^3$ vertices, each of them possessing
six qubits (except the ones lying on the sides of the cube), on which arbitrary
quantum operations can be applied perfectly. Nearest
neighbors share one partially entangled state of the form
\begin{multline}
	\rho = (1-\eps)^2 \ketbra{\Phi^+}{\Phi^+} +
			\eps(1-\eps) \ketbra{\Psi^+}{\Psi^+} \\
			+\eps(1-\eps) \ketbra{\Phi^-}{\Phi^-} +
			\eps^2 \ketbra{\Psi^-}{\Psi^-}.
	\label{eq:rho}
\end{multline}
Such a state can be realized as follows: a station prepares locally
a maximally entangled pair of qubits, and sends one of them to a neighbor.
In the quantum channel, the traveling qubit undergoes random and independent
bit-flip and phase errors with probability $0<\eps<0.5$. This channel describes a specific
physical process, but the generality of $\rho$ is in reality complete.
In fact, we show in App.~\ref{app:rho} that any entangled state of two qubits
can be brought to this form by local quantum operations and classical communication.
Finally, all classical processes (communication and computation) are assumed to
take much less time than any quantum operation.\\

\paragraph*{Remark 1.}
Physical implementations of three-dimensional lattices have been proposed
in the context of quantum information processing and distributed quantum computation
\cite{BCJD99,IM09}. For practical reasons, however, it may be advantageous
to realize the proposed construction in two dimensions, using a ``slice-by-slice''
generation similar to the techniques developed in \cite{RH07}. In that case
note that the time required to run the protocol scales linearly with $N$.

\paragraph*{Remark 2.}
Recently, ideas of percolation theory have been applied successfully to the
case of mixed states of rank two \cite{BDJ09}. In addition to the fact that the techniques
are very different, our study is not restricted to amplitude damping channels,
but considers full-rank mixed states which are robust against any small perturbations.
In fact our protocol still works if dependent bit-flip and phase errors are present in
the connections.

\section{A mapping to noisy cluster states}
It was shown in \cite{PJS+08} how to create and propagate a large Greenberger-Horne-Zeilinger (GHZ) state
in a noisy square lattice. This state is robust against bit flips if their
rate is not too high but is very fragile against phase errors. Any
of them indeed destroys the coherence of the GHZ state. Therefore, an encoding of the qubits is required,
which leads to a logarithmic scaling of the physical
resources per node. Since we are looking for a fidelity
threshold, we want to create a large state that has the ability
to correct both bit-flip and phase errors. Cluster states thus arise as a natural choice.
In fact they have been shown
to possess an intrinsic capability of error correction, so that long-range entanglement
between two faces of an infinite noisy cubic cluster state is indeed possible \cite{RBH05}.
Our protocol is based on this construction, with two radical differences, however:
first, the settings are distinct, and second, we allow only local quantum
operations on \textit{all} the nodes.

\begin{figure}
	\begin{center}
    	\includegraphics[width=7cm]{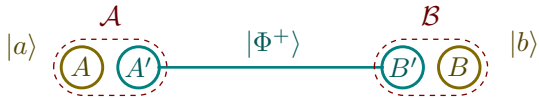}
    	\caption{(Color online) Non-local control phase on two qubits $A$
    	and $B$, with the help of a Bell pair $\ket{\Phi^+}_{A'B'}$.}
    \label{fig:cphase}
	\end{center}
\end{figure}

A cluster state, which is an instance of graph states,
can be constructed by inserting a qubit in the state $\ket{+}$ at each vertex of the
graph and by applying a control phase between all neighboring pairs \cite{HDR+06}.
In our setting we cannot perform these control phases since they are non-local
quantum operations, but we can add an ancillary qubit and perform joint measurements
at each node such that the resulting state is a cluster state.
This method has been described in \cite{VC04} in the case
of perfect links, which can be interpreted as the virtual components of a large
valence-bond state, and has been generalized to imperfect connections in \cite{RBH05}.
Nevertheless, let us describe here an explicit (and slightly
different) construction, mainly for completeness sake but also for relating precisely
the error rate in the quantum networks with the one in the noisy cluster state.

At each node, we add a qubit $\ket{+}$ and use the noisy links $\rho$ to indirectly
perform the control phases. Let us first describe how this
is achieved if all connections are perfect, \ie their qubits are in the state $\ket{\Phi^+}$.
We consider two nodes of the lattice, with two qubits $A$ and $B$ in the states
$\ket{a}=a_0\ket{0}+a_1\ket{1}$ and $\ket{b}=b_0\ket{0}+b_1\ket{1}$, and a connection $\ket{\Phi^+}$ between
two qubits $A'$ and $B'$, see Fig.~\ref{fig:cphase}.
We start by applying, on the qubits of the first node, the measurement operators
\begin{equation}
\begin{split}
	\cA_0 &= \ket{0}_A\bra{00}_{AA'}+\ket{1}_A\bra{11}_{AA'},\\
	\cA_1 &= \ket{0}_A\bra{01}_{AA'}+\ket{1}_A\bra{10}_{AA'},
\end{split}
\end{equation}
with $\sum_{i=0}^1\cA_i^{\dagger}\cA_i^{\vphantom{\dagger}}=\id_4$, which
are followed by a bit flip $X$ on $B'$ if the outcome is $\cA_1$.
The resulting state on $A$ and $B'$ reads $a_0\ket{00}+a_1\ket{11}$. We then apply
the second measurement
\begin{equation}
\begin{split}
	\cB_0 &= \ket{0}_B\bra{+0}_{B'B}+\ket{1}_A\bra{-1}_{B'B},\\
	\cB_1 &= \ket{0}_B\bra{-0}_{B'B}+\ket{1}_A\bra{+1}_{B'B},
\end{split}
\end{equation}
followed by the matrix $Z$ on $A$ if we get $\cB_1$ as outcome. Finally,
$A$ and $B$ are left in the (entangled) state
\begin{equation*}
	\ket{C_{ab}} = a_0b_0\ket{00}+a_0b_1\ket{01}+a_1b_0\ket{10}-a_1b_1\ket{11},
\end{equation*}
which is the result of a control phase between $\ket{a}$ and $\ket{b}$. Clearly,
if $\ket{a}=\ket{b}=\ket{+}$, the state $\ket{C_{ab}}$ is the cluster state on
two qubits. Now, let us determine which errors occur if we blindly perform
the very same operations but using another Bell state. It is straightforward
to compute the results of these operations if one uses $\ket{\Psi^+}$,
$\ket{\Phi^-}$, or $\ket{\Psi^-}$ for the connections:
one gets $\id_2 \otimes Z \ket{C_{ab}}$, $Z \otimes \id_2 \ket{C_{ab}}$, or
$Z \otimes Z \ket{C_{ab}}$, respectively.
Since the matrices $Z$ commute with the control phases, it follows that errors
do not propagate while constructing a (noisy) cluster state $\rho_{\text{CS}}$ from the cubic
quantum network. Moreover, because of the specific choice of coefficients in Eq.~(\ref{eq:rho}),
$Z$ errors appear independently at the nodes. Since a node
of the lattice has degree six at most, and two $Z$ errors cancel each other,
the vertices of the resulting cluster state suffer an error with a probability
at most equal to
\begin{equation}
	p = \sum_{i=0}^2 \binom{6}{2i+1}\eps^{2i+1}(1-\eps)^{5-2i}.
	\label{eq:perror}
\end{equation}
This expression reduces to $p \approx 6\eps$ in the regime of small error rates.
Therefore, we are exactly in the setting of \cite{RBH05}, where thermal
fluctuations in the cluster state induce independent local $Z$ errors with rate $p$.\\

\section{Long-range entanglement in noisy cluster states}
In this section, we mainly follow the construction and the notation proposed in \cite{RBH05},
namely, the measurement of the qubits of $\rho_{\text{CS}}$ according to a
specific pattern of local bases. The outcomes of the measurements
are random, but the choice of the bases establishes some parity constraints on them.
Any violation of these constraints indicates an error, and a classical processing of
all collected ``syndromes'' allows one to reliably identify the typical errors. This
correction works perfectly for small error rates, but it breaks down at $p_c\approx3.3\%$ \cite{OAIM04}.
The difference between the present method and that given in \cite{RBH05} is that no non-local quantum operation
is allowed. This obliges us to design a more elaborated error correction, leading to
a different type of long-distance entanglement. In fact we are not going to create a pure
and perfect Bell pair of logical qubits, but rather a mixture of two entangled
physical qubits.

\subsection{Measurement pattern and long-distance quantum correlations}

\begin{figure}
	\begin{center}
    	\includegraphics[width=\linewidth]{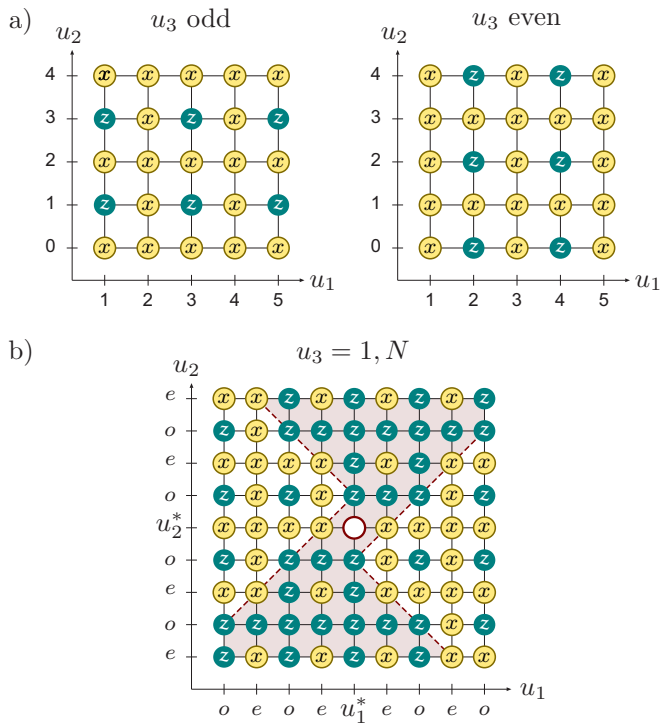}
    	\caption{(Color online) a) Bases in which qubits
    	    in the bulk of the cluster state are measured.
    		b) Slightly different measurement pattern for the faces $\cL$ and $\cR$:
    		the central qubit is kept intact, and all qubits that lie
    		in the shaded area are measured in the $Z$ basis, except the
    		ones with coordinates $(e,e,1)$ or $(e,e,N)$, which are measured
    		in the $X$ basis.}
    \label{fig:measure}
	\end{center}
\end{figure}

Let us define a finite three-dimensional cluster state on the cube
\[
	\cC = \{u=(u_1,u_2,u_3): 1\leq u_1,u_2+1,u_3\leq N\},
\]
and select two qubits $A$ and $B$ centered in two opposite faces $\cL$
and $\cR$. The coordinates of these qubits are $(u_1^*,u_2^*,1)$ and $(u_1^*,u_2^*,N)$,
with $u_1^*=u_2^*+1=(N+1)/2$. For a reason that will soon become clear,
we consider lattices of size $N\equiv 1$ (mod 4), so that $u_1^*$ is odd
and $u_2^*$ even. Let us also introduce two disjoint sublattices $T_o$ and
$T_e$ with double spacing, where $o$ and $e$ stand for odd and even.
Their vertices are
\begin{align*}
	V(T_o) &= \{u=(o,o,o)\} \subset \cC,\\
	V(T_e) &= \{u=(e,e,e)\} \subset \cC,
\end{align*}
and their edges are given by the sets
\begin{align*}
	E(T_o) &= \{u=(o,o,e),(o,e,o),(e,o,o)\} \subset \cC,\\
	E(T_e) &= \{u=(e,e,o),(e,o,e),(o,e,e)\} \subset \cC.
\end{align*}
We also define the planes
\begin{align*}
	T_X^{(u_2)} = \{u=(o,u_2,o)\} \subset T_o,\\
	T_Z^{(u_1)} = \{u=(u_1,e,e)\} \subset T_e,
\end{align*}
and denote by $T_X^*$ and $T_Z^*$ the planes that contain $A$ and $B$.
These planes will be used to derive the Bell correlations of the future long-distance
entangled state $\rs$ (we first consider that no error occurs, and then
extend the results to noisy cluster states).
Qubits that belong to the vertices of $T_o$ and $T_e$ are measured in the $Z$ basis,
while all other qubits are measured in the $X$ basis. There are, however, some
exceptions in $\cL$ and $\cR$ (see Fig.~\ref{fig:measure}): First, the central
qubit is not measured, since it will be part of the long-distance entangled state.
Second, qubits with coordinates $u_1=u_1^*$ are measured in the $Z$ basis in order to create the right
quantum correlations, as explained in the following paragraph. Finally, we
measure in the $Z$ basis all qubits whose first two coordinates are $(e,o)$ or $(o,e)$
and which lie in the shaded areas; these outcomes will be important
for the error correction.

To compute the effect of the
measurements on the quantum correlations between $A$ and $B$,
we use the fact that a perfect cluster state $\cs$ obeys the eigenvalue
equation $K_u \cs = \cs$ for all $u\in\cC$, where $K_u$ is the
stabilizer
\begin{equation}
	K_u = X_u \prod_{v\,\in\,\cN(u)} Z_v,
\end{equation}
with $\cN(u)$ the neighborhood of $u$. If we let the products of stabilizers
$\prod_{u\,\in\,T_X^*}K_u$ and $\prod_{u\,\in\,T_Z^*}K_u$
act on the cluster state,
we find that $A$ and $B$ are indeed maximally entangled:
\begin{align}
	X_A X_B\rs &= \lambda_X\rs,\nonumber\\
	Z_A Z_B\rs &= \lambda_Z\rs,
	\label{eq:correl}
\end{align}
with $\lambda_X,\lambda_Z\in\{-1,+1\}$. The eigenvalues $\lambda_{X,Z}$ are calculated
from the measurement outcomes $x$ and $z$:
\begin{align}
	\lambda_X &= \prod_{u\,\in\,\Omega_X^{(z)}}z_u
					\prod_{u\,\in\,\Omega_X^{(x)}}x_u,\nonumber\\
	\lambda_Z &= \prod_{u\,\in\,\Omega_Z^{(z)}}z_u
					\prod_{u\,\in\,\Omega_Z^{(x)}}x_u,
\end{align}
where
$\Omega_X^{(x)} = T_X^*\setminus\{A,B\}$,
$\Omega_Z^{(x)} = T_Z^*$,
$\Omega_X^{(z)} = T_X^{(u_2^*+ 1)} \cup T_X^{(u_2^*- 1)}$, and
$\Omega_Z^{(z)} = T_Z^{(u_1^*+ 1)} \cup T_Z^{(u_1^*- 1)} \cup \{(u_1^*,e,1),(u_1^*,e,N)\} \setminus \{A,B\}$.

\subsection{Error correction}
\begin{figure}
	\begin{center}
    	\includegraphics[height=3.9cm]{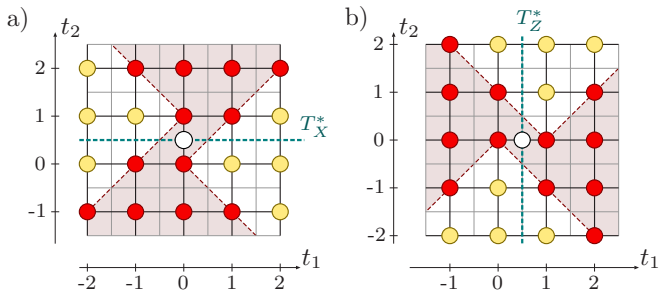}
    	\caption{(Color online) a) In red (dark disks), missing syndromes on the vertices
    		of $T_o$, in $\cL$ and $\cR$, corresponding
    		to the measurement pattern depicted in Fig.~\ref{fig:measure}c.
    		Rough faces lie on the top and the bottom of this lattice.
    		A new coordinate system $(t_1,t_2,t_3)$ is introduced for the vertices of the sublattice.
    		b) Same considerations for the $Z$ correlation: 
    		the missing syndromes create additional rough surfaces in $T_e$.}
    \label{fig:syndromes}
	\end{center}
\end{figure}

As already mentioned, measurement outcomes are random but not independent. It is
thus possible to assign to most vertices $u_i\in T_i$, with $i=o$ or $e$, the parity syndrome 
\begin{equation}
	s(u_i) = \prod_{v\,\in\,\cN(u_i)}x_v \prod_{w\,\in\,\cN_i(u_i)} z_w,
\end{equation}
where $\cN_i(u_i)$ designates the neighborhood of $u_i$ in $T_i$.
Since this equation arises from a product of stabilizers, $\prod_{v\in\cN_i(u_i)}K_v$,
we have that $s(u_i)=1$ if no error occurs on the qubits of $\cN_i(u_i)$. The key point
of the construction is that a $Z$ error on any edge of $T_i$ changes the sign
of the two syndromes at its extremities. This is due to the fact that $Z$ errors do not
commute with $X$ measurements, while outcomes $z$ are not affected by
them. The sublattices are treated separately, but in a similar way. We
refer the reader to \cite{RBH05,DKLP02} for a detailed discussion of the error
recovery or to App.~\ref{app:correction} for the basics to understand our protocol.
In contrast with \cite{RBH05}, and apart from the rough faces present in any surface code,
we also suffer a lack of syndrome information in $\cL$ and $\cR$.
We cannot have a perfect and complete syndrome pattern for both $T_o$ and $T_e$
in these faces; for this to happen one should be able to measure both $x$ and $z$
eigenvalues of the concerned qubits, which is impossible, or apply non-local quantum
operations, which we do not allow. Actually, useful long-distance quantum correlations can
still be created if one performs the measurements depicted in Fig.~\ref{fig:measure}b:
half outcomes are used to gain information on $T_o$, and symmetrically for $T_e$,
see Figs.~\ref{fig:syndromes} and \ref{fig:paths}.\\

As an example of the effect of the unknown syndromes in $\cL$,
let us consider that an error occurred on the center qubit $A$, and that all other qubits
did not suffer any error. Since we do not know the syndromes of $T_o$ that lie
directly below and above $T_X^*$, we are not able to restore the $X$ correlation.
This occurs with probability $p_X = p+\mathcal{O}(p^2)$. From
this fact, one finds that the final state on $A$ and $B$ is a mixed state of the form
\begin{multline}
	\rho_{AB} = F_X F_Z \ketbra{\Phi^+}{\Phi^+} +
			p_X F_Z \ketbra{\Psi^+}{\Psi^+} \\
			+F_X p_Z \ketbra{\Phi^-}{\Phi^-} +
			p_X p_Z \ketbra{\Psi^-}{\Psi^-},
\end{multline}
with $F_X=1-p_X$ and $F_Z=1-p_Z$.
This state is known to be distillable, and thus useful from a quantum information perspective,
whenever its fidelity $F_{AB}\equiv F_X F_Z$ is larger that one-half \cite{DEJ+96}.
This can be achieved when the error rate $p$ is smaller than a threshold $p^*$.
In the next paragraphs, we first prove a lower bound on this value, $p^*\gtrsim1.17\times10^{-3}$,
and then present numerical results, showing that the real threshold is indeed much larger:
$p^*\gtrsim \ps\%$.

\subsubsection{Correlation loss due to the missing syndromes in $T_o$}
\begin{figure}
	\begin{center}
    	\includegraphics[height=3.9cm]{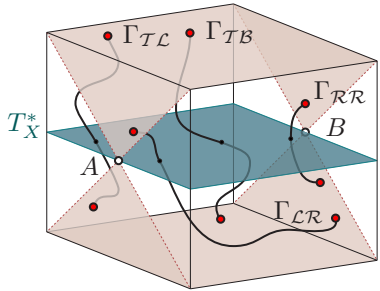}
    	\caption{(Color online) Some paths of errors that have a non-trivial
    	effect on the long-distance entanglement:
    	any path stretching from one shaded area to another, and crossing the plane
    	$T_X^*$ an odd number of times, degrades the $X$ correlation between $A$
    	and $B$. The shaded areas, which partially wrap the cube, are of two
    	types: the top and bottom ones are the usual rough surfaces present in
    	(three-dimensional) surface codes. The left and right shaded areas
    	represent the unknown syndromes of $T_o$, see Fig.~\ref{fig:syndromes}a.
    	The situation for $T_e$ is very similar (the picture is rotated by
    	$90^\circ$ around the $AB$ axis), with the difference that all shaded
    	areas are rough surfaces in this case.}
    \label{fig:paths}
	\end{center}
\end{figure}

Paths of errors, which we generically denote by $\Gamma$, have a non-trivial
effect on the $X$ correlation if they cross the plane $T_X^*$ an odd number of
times, as depicted in Fig.~\ref{fig:paths}. Moreover, the number $l$ of errors
which actually occur on a path $\Gamma$ is at least $L/2$, where $L$ denotes its length.
This is the case because our error correction always leads to a minimum
pairing of the syndromes $s=-1$. We now follow Chap.~V in \cite{DKLP02}
to find an upper bound on the probability $p_X$ of inferring the wrong quantum correlation:
\begin{multline}
	p_X \leq
		2\sum_{\GLL}\prob(\GLL) + \sum_{\GLR}\prob(\GLR)\\
		+\sum_{\GTB}\prob(\GTB) + 4\sum_{\GTL}\prob(\GTL),
	\label{eq:boundpX}
\end{multline}
where $\cB$ and $\cT$ stand for the bottom and top faces. Note that we already
took into account the symmetries of the problem in this expression.
For convenience, let us now introduce a new coordinate system $(t_1,t_2,t_3)$ for
the vertices of $T_o$, such that $-N_o\leq t_1\leq N_o$, $-N_o < t_2\leq N_o$, and
$0\leq t_3\leq 2 N_o$, with $N_o = (N-1)/4$, see also Fig.~\ref{fig:syndromes}.
In this coordinate system, paths of errors $\GTL$ travel a distance $L\geq N_o$
and can start from $N_o^2$ missing syndromes in $T_o$ (lower triangle in $\cL$).
Because for each vertex there are, in a cubic lattice, at most $5^L/2$ self-avoiding
walks pointing upward, we find that the last term of Eq.~(\ref{eq:boundpX}) is upper bounded by
\begin{equation*}
	\sum_{\GTL}\prob(\GTL) \leq N_o^2 \sum_{L\geq N_o} \frac{5^L}{2}
	\sum_{l=\lceil L/2\rceil}^L \binom{L}{l}\,p^l\,(1-p)^{L-l},
\end{equation*}
where $\lceil L/2\rceil$ denotes the smallest integer not less than $L/2$.
The sum over $l$, together with the binomial coefficients, counts all
possible paths of errors that appear in a given walk. One can check that the bound
tends to 0 in the limit $N_o\rightarrow\infty$
if $10\sqrt{p(1-p)}<1$, \ie if $p\lesssim 1\%$. The same result holds for
the paths $\GLR$ and $\GTB$; note that this value is about three times smaller
than the real critical point $p_c\approx3.3\%$. Similar considerations for the
paths $\GLL$ finally yield, for $p<1\%$,
\begin{align}
	p_X &\leq 2\sum_{\GLL}\prob(\GLL)\nonumber\\
		&\leq 2 \sum_{t_2\geq1} 2\,t_2 \sum_{L\geq t_2}
			\frac{5^L}{2} \sum_{l \geq \lceil L/2\rceil} \binom{L}{l}\,p^l\,(1-p)^{L-l}.
	\label{eq:pX}
\end{align}
This bound never tends to zero, but still converges if $p$ is small enough.
Before computing a threshold for $F_{AB}$, however, we first have to consider
the errors made in the other sublattice. 

\subsubsection{Loss of correlation in $T_e$ and fidelity of the final state}
\begin{figure}
	\begin{center}
    	\includegraphics[width=.98\linewidth]{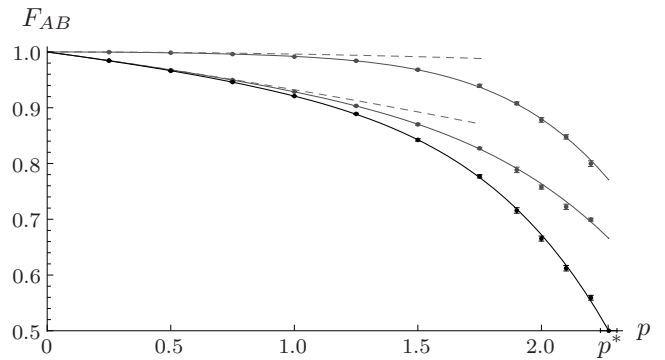}
    	\caption{Fidelity $F_{AB}=F_X F_Z$ of the long-distance entangled
    	state $\rho_{AB}$ as a function of the error rate $p$ (in percent). The critical
        value $F^*=0.5$ is reached at $p^* = \ps \pm \dps\,[\%]$.
    	The upper curve represents the probability of success $F_Z$ of the error
    	correction in $T_e$, while the middle one is the function $F_X$.
    	Corresponding series expansions for $p\ll1$ are plotted with dashed lines.}
    \label{fig:FXZ}
	\end{center}
\end{figure}

The situation for the $Z$ correlation is very similar to the previous
case, since the measurement pattern is symmetric. Nonetheless there is a small difference:
the missing syndromes do not lie on the vertices of the sublattice, but rather on its
outer edges $\cL$ and $\cR$. This creates additional parts of rough faces in $T_e$.
It follows that the corresponding paths of errors have their first and last edges pointing
in the $t_3$ direction, so that a slightly better bound for the error $p_Z$ can
be derived:
\begin{equation}
	p_Z \leq 2 \sum_{t_1\geq1} 2\,t_1 \myNeg\sum_{L\geq t_1+2}\myNeg
			\frac{5^{L-2}}{2} \myNeg\sum_{l \geq \lceil L/2\rceil} \binom{L}{l}\,p^l\,(1-p)^{L-l}.
	\label{eq:pZ}
\end{equation}
Combining the two bounds on $p_X$ and $p_Z$ we find
\begin{equation}
	p\lesssim 1.17 \times 10^{-3} \,\Rightarrow\, F_{AB}>1/2,
\end{equation}
which corresponds, via Eq.~(\ref{eq:perror}), to an error rate $\eps \lesssim 1.95 \times 10^{-4}$ in
the initial connections. This value is quite small, mainly because our counting
of paths of errors is very crude. Note that there are only few such paths of small length,
and therefore this analytical bound could be increased by carefully computing
its smallest orders in $p$. At this point, however, we prefer to turn
to Monte Carlo simulations to find a much better estimate of the error threshold.
We refer the reader to App.~\ref{app:MC} for a description of the
algorithm; in particular we propose an intuitive and efficient method, even if
not optimal, to infer the value of the missing syndromes.
The result of these simulations is plotted in Fig.~\ref{fig:FXZ}: long-distance
entanglement is achieved for error rates smaller than
\begin{equation}
	p^* \approx \ps \%,
\end{equation}
\ie $\eps^*\approx \es \times 10^{-3}$ for the original lattice. Before
concluding, let us comment on these thresholds:
\begin{itemize}
	\item The values of the unknown syndromes are not optimally inferred in our algorithm,
	and therefore a higher value of $p^*$ may be found. However, it is clear that
	it cannot exceed the critical error rate $p_c\approx3.3\%$.
	\item One could get a higher threshold $\eps^*$ by directly computing
	$F_{AB}$ as a function of $\eps$. In fact, errors in the faces $\mathcal{L}$ and
	$\mathcal{R}$ do not appear with probability $p\approx6\eps$, but
	only with probability $p\approx5\eps$.
	\item Our measurement pattern puts $T_o$ and $T_e$ on the same footing
	(Figs.~\ref{fig:measure} and \ref{fig:syndromes}), but it could be
	profitable to get more information on the unknown syndromes of $T_o$
	since the $X$ correlation is more sensitive to errors.
	\item Finally, as suggested in \cite[Rem. 2]{RBH05}, lattices of size
	$\log(N)\times\log(N)\times N$ may also be appropriate for
	generating long-distance entanglement. This result also holds
	in our setting, because additional errors only appear in the
	faces $\cL$ and $\cR$ and not in the bulk of the lattice.
\end{itemize}

\section{Conclusion}
We have investigated the problem of generating long-distance entanglement in
noisy quantum networks. We have focused on three-dimensional regular lattices,
whose edges are full-rank mixed states of two qubits. We have proven that
entanglement can be established between two infinitely distant qubits
if the fidelity of the connections is large enough. Our protocol starts by
transforming the quantum network into a thermal cluster state. Then, all but two distant qubits
are measured according to a specific pattern of local bases, and a syndrome-based
error correction is performed. The error recovery is very similar to the one
used for planar codes, with the difference being that our setting does not allow one to
get complete information on the syndromes. Nevertheless, useful quantum correlations can
be created between the two unmeasured qubits if the error rate is
smaller than a critical value. We have given both an analytical lower bound on this value
and a numerical estimation (about 2\%) based on Monte Carlo simulations.\\

In conclusion, we have shown that a constant overhead of local resources is sufficient to
achieve long-distance communication in quantum networks. This contrasts with previous one- or
two-dimensional strategies, in which the physical resources per station increase with the
distance. Our protocol requires perfect local quantum operations,
which is somehow justified by the fact that most errors occur while sending
quantum information between stations. It would nevertheless be of fundamental interest to
design fault-tolerant protocols, in two or three dimensions, for which the
overhead of local resources is as small as possible.

\begin{acknowledgments}
The author thanks Ignacio Cirac and Antonio Ac\'in for initiating the project
and for useful discussions.
This work has been supported by the QCCC program of the Elite Network of Bavaria.
\end{acknowledgments}

\appendix
\numberwithin{equation}{section}

\section{Elementary entangled pairs of qubits}
\label{app:rho}
We show in this appendix that there is no loss of generality in choosing the
elementary links to be described by Eq.~(\ref{eq:rho}). First, it is well known
that any two-qubit entangled state can be brought to the rotationally
symmetric mixture
\begin{multline}
	W_F = F \ketbra{\Phi^+}{\Phi^+} +
			\frac{1-F}{3} \ketbra{\Psi^+}{\Psi^+}  \\
			+\frac{1-F}{3} \ketbra{\Phi^-}{\Phi^-} +
			\frac{1-F}{3} \ketbra{\Psi^-}{\Psi^-}.
\end{multline} 
In fact, this is achieved by applying random bilateral rotations, locally, to each qubit.
This state is called a Werner state \cite{W89} and has the same fidelity
$F$ as the initial state from which it derives. Let us denote $W_F$
by its components in the Bell basis, $W_F = (F,\frac{1-F}{3},
\frac{1-F}{3},\frac{1-F}{3})$, and define the two unitaries
\begin{equation}
	H  = \frac{1}{\sqrt{2}}\begin{pmatrix}1&1\\1&-1\end{pmatrix}\quad\text{and}\quad
	H' = \frac{1}{\sqrt{2}}\begin{pmatrix}i&1\\1&i\end{pmatrix}.
\end{equation}
One can check that the separable operation $H \otimes H$ applied on a Bell-diagonal state
switches its coefficients $\Phi^-$ and $\Psi^+$, while the coefficients $\Phi^+$
and $\Psi^-$ are unaltered. A similar result holds for
$H' \otimes H'$, which only switches the components $\Phi^+$ and $\Psi^+$. Suppose now
that an entangled pair $W_F$, with $F=1-3\,\eps^2$, has been created
between two neighboring nodes. This already sets the coefficient $\Psi^-$ to
the desired value $\eps^2$. Then, apply $H' \otimes H'$ with probability
$p=\frac{2\eps}{1+2\eps}$, and $\id \otimes \id$ with probability $1-p$ on $W_F$. This
leads to the state $\big((1-\eps)^2,\eps(2-3\eps),\eps^2,\eps^2\big)$. Finally,
repeat the operation by applying $H \otimes H$ with probability one-half.
Both coefficients $\Psi^+$ and $\Phi^-$ are set to $\eps(1-\eps)$, which
proves that the state $\rho$ given in Eq.~(\ref{eq:rho}) is indeed general.\\

\section{Basics of syndrome-based error correction}
\label{app:correction}
Let us consider the sublattices $T_o$ and $T_e$ described in the text, in which
$Z$ errors occur independently on each edge with probability $p$, and assign to each
vertex the syndrome $s=+1$ if it is connected to an even number of erroneous edges,
and $s=-1$ otherwise.
In the case of perfect and complete syndrome information, one knows exactly where
all paths of errors start and end: this occurs at syndromes $s=-1$. In the regime of small error
rate $p$, it turns out that the best error recovery strategy is to pair these syndromes such
that the total length of all pairings is minimized. Then, one connects any two
paired syndromes by a path of minimum length, and artificially introduces $Z$ ``errors''
along these paths. This creates loops of errors in the cluster state, which, however,
do not cause any damage to the long-distance quantum correlations. In fact these
loops either do not intersect the planes $T_X^*$ and $T_Z^*$ or cross them twice,
and consequently do not modify the eigenvalues in Eq.~(\ref{eq:correl}).

Problems arise because some syndromes are unknown. For instance, consider the edges that have
only one extremity in $V(T_o)$ or $V(T_e)$: their coordinates are $(o,0,o)$ and
$(o,N-1,o)$ in $T_o$, and $(1,e,e)$ and $(N,e,e)$ in $T_e$, see Fig.~\ref{fig:syndromes}.
These are the rough faces described in \cite{RBH05}, and errors on these edges change the sign of
only one syndrome (and not two) in the corresponding sublattice. An equivalent viewpoint
is that both extremities of these edges indeed belong to $T_o$ or $T_e$, but we
do not have access to their outer syndrome. The consequence of this lack of information
is that some paths of errors $\Gamma$ are not closed anymore, but rather originate from a
missing syndrome and terminate at another. Typically, these open paths enter only
superficially the lattice if the error rate $p$ is small, but they start stretching
from one side to another as soon as $p$ exceeds the value $p_c\approx3.3\%$.
In the latter case, paths of errors can cross an odd number of times the planes
of correlations, which results in a complete loss of long-distance entanglement
in the limit $N\rightarrow\infty$.\\

\section{Monte Carlo simulations}
\label{app:MC}
We now describe the algorithm used for computing the data of Fig.~\ref{fig:FXZ}.
The correction procedure is very similar for the two sublattices $T_o$ and $T_e$, and consists of two main parts.
First, given a lattice with random errors, we infer the value of the syndromes
for which we have no information. Second, we proceed with the usual error recovery.
The program outputs 1 if the correction is successful, and 0 otherwise.

\begin{figure}
	\begin{center}
    	\includegraphics[height=4cm]{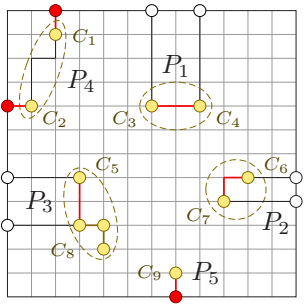}
    	\caption{(Color online)
    	An example of how unknown syndromes, which here lie on the boundary,
    	are assigned the value $\pm1$. We first find all odd-size clusters $C_i$ of
    	syndromes $-1$, which are drawn in light yellow (light gray), and pair them
    	by increasing distance. A cluster may be left alone; in that case we add it in the end
    	of the sorted list of cluster pairs $P_i$. Then, for each $P_i$,
    	we check if it is favorable to add two new syndromes $-1$. Here, this
    	is the case only for $P_4$ and $P_5$.}
    \label{fig:typicalcluster}
	\end{center}
\end{figure}

\subsubsection{Inferring the missing syndromes}
We propose a very simple way of assigning the value $+1$ or $-1$ to the missing
syndromes, so that a good approximation of the optimal configuration is found.
To that end, it is helpful to consider a typical realization of a noisy cluster
state in the regime of small error rates, as depicted in Fig.~\ref{fig:typicalcluster}.
Our algorithm reads
\begin{itemize}
	\item Initialize all unknown syndromes to $+1$.
	\item Using nearest-neighbor site percolation, find all clusters $C_i$
	of syndromes $-1$. Keep only the clusters of odd size,
	and for each compute the minimum distance $d_i$ to a closest
	unknown syndrome $s_i$.
	\footnote{Several such syndromes may exist; choose the one
	that lie in the plane parallel to $T_X^*$ or $T_Z^*$ that contains $C_i$.
	This avoids unnecessary crossings of the plane of correlation.}
	Let $n$ denote the number of such clusters.
	Note that we do not consider clusters
	of even size, since good pairings can be found for them, individually.
	\item Calculate the distance $d_{ij}$ between all pairs of clusters $C_i$
	and $C_j$. This distance could be the length of the shortest path from $C_i$
	to $C_j$, but in practice it is much easier to calculate the distance between their
	``centers of mass.''
	\item Find $C_a$ and $C_b$ such that $d_{ab} = \min\{d_{ij}\}$, and
	create a pair $\mathcal{P}_1 = \{C_a, C_b\}$. Remove $C_a$ and $C_b$
	from the list of clusters, and repeat the procedure until no cluster
	is left. In case of odd $n$, add an extra ``pair'' $\{C_i,C_i\}$ for the
	remain cluster,
	with $d_{ii}=\infty$. This creates the list $\{P_1,\ldots,P_{n/2}\}$.
	\item For each $P_k = \{C_i,C_j\}$, check if $d_{ij}>d_i+d_j$.
	If this inequality holds, inverse the value of the corresponding missing
	syndromes: $s_i\leftarrow (-s_i)$ and $s_j\leftarrow (-s_j)$.
\end{itemize}
The proposed algorithm is optimal in the regime of very dilute errors, but this
is not true for high error rates anymore (even if results are good for
all $p$). Note that there exist optimal algorithms which are based on minimal perfect
matchings in weighted graphs and are run in a polynomial time (see \cite{WFSH09}, Chap.~4).
For three-dimensional lattices, however, the number of edges in these
graphs scales as $\mathcal{O}(N^6)$ [they are nearly complete graphs on $\mathcal{O}(N^3)$ vertices],
and therefore these algorithms are not so efficient in practice.
Nevertheless, it would be very interesting to implement an optimal algorithm
and decide whether the unknown syndromes are responsible for the threshold,
or whether the equality $p^*=p_c$ holds.\\

\subsubsection{Error recovery}
\begin{figure}
	\begin{center}
    	\includegraphics[width=.98\linewidth]{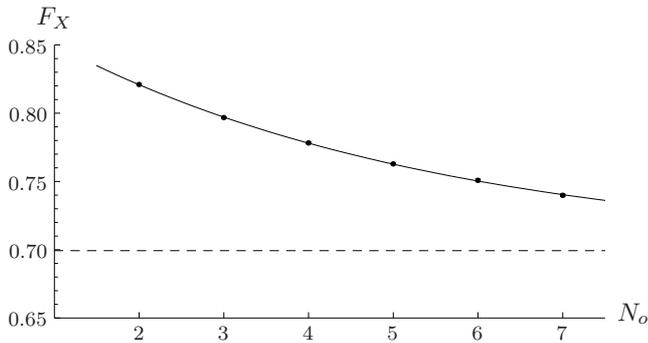}
    	\caption{Fidelity $F_X$ in the limit $N_o\rightarrow\infty$, for a
    	fixed error rate $p$. We use the function $F_X^{(\infty)} + a\,e^{-b\,N_o}$
    	to fit the data computed from lattices consisting of
    	$(2N_o+1)\times(2N_o)\times(2N_o+1)$ nodes. At least $10^5$ simulations
    	have been run for each value of $N_o\in\{2,3,\ldots,7\}$.
    	In this example the error rate is $p=2.2\%$, and the fit yields
    	$F_X^{(\infty)} = \fx \pm \dfx$.}
    \label{fig:extrapolate}
	\end{center}
\end{figure}

We use the well-known and efficient algorithm described by J. Edmonds in \cite{E65B}
to find an optimal pairing of the syndromes $-1$. The error correction is
successful if the parity of paths of errors crossing the plane of correlation
is even. Simulations of the error corrections have been performed for various
lattice sizes (up to $\approx 15^3$ nodes), and for both $X$ and $Z$ correlations. 
The extrapolation to infinite lattices is done by fitting the data with an
exponential function, see Fig.~\ref{fig:extrapolate}. Results are
plotted in Fig.~\ref{fig:FXZ}.\\

Finally, let us present evidence that our algorithm gives correct and
optimal results in the regime of small error rates. Considering the series expansions of $p_X$
at first order in $p$, one sees that only three edges of $\mathcal{L}$ may
degrade the $X$ correlation: these are the bonds in $T_o$ that cross $T_X^*$
and whose first coordinate $t_1$ belongs to $\{-1,0,1\}$, see Fig.~\ref{fig:syndromes}a.
At second order, one can check that the probability to infer the wrong $X$ correlation due to the missing
syndromes in $\mathcal{L}$ is $p_X^{\mathcal{L}}=3p+48p^2$. Therefore, by
symmetry, the fidelity $F_X=1-p_X^{\mathcal{L}}(1-p_X^{\mathcal{R}})-
p_X^{\mathcal{R}}(1-p_X^{\mathcal{L}})$ reads
\begin{equation}
	F_X = 1-6p-78p^2+\mathcal{O}(p^3).
	\label{eq:expandFX}
\end{equation}
It is easy to see that a single error in $T_e$ cannot damage the $Z$ correlation,
and a careful counting of configurations with two errors yields $p_Z^{\mathcal{L}}=19p^2$.
Consequently we find
 \begin{equation}
	F_Z = 1-38p^2+\mathcal{O}(p^3).
	\label{eq:expandFZ}
\end{equation}
These two series expansions are plotted (dashed lines) in Fig.~\ref{fig:FXZ}: they
agree perfectly with the results of the Monte Carlo simulations and thus
validate our algorithm.\\

\bibliography{article}

\end{document}